# Core-valence double ionization of $SF_6$ involving S2$p$, F1$s$ and S1$s$ inner shells

**Veronica Daver Ideböhn[1], Daniel M. Pereira[2], Lucas M. Cornetta[2], Richard J. Squibb[1], Andreas Hult Roos[1], Emelie Olsson[1], Måns Wallner[1], Marco Parriani[1,3], Nihar Ranjan Behera[1], Gunnar Öhrwall[4], Florian Trinter[5], John H.D. Eland[6], Hans Ågren[7], and Raimund Feifel[1,+]**

[1]University of Gothenburg, Department of Physics, Origovägen 6B, 412 58 Gothenburg, Sweden
[2]Instituto de Física, Universidade de São Paulo, Rua do Matão 1731, São Paulo, 05508-090 Brasil
[3]University of Perugia, Department of Civil and Environmental Engineering, Via G. Duranti 93, 06125 Perugia, Italy
[4]MAX IV Laboratory, Lund University, Box 118, 221 00 Lund, Sweden
[5]Molecular Physics, Fritz-Haber-Institut der Max-Planck-Gesellschaft, Faradayweg 4-6, 14195 Berlin, Germany
[6]Oxford University, Department of Chemistry, Physical and Theoretical Chemistry Laboratory, South Parks Road, Oxford OX1 3QZ, United Kingdom
[7]Department of Physics and Astronomy, Uppsala University, Box 516, 751 20 Uppsala, Sweden
[+]raimund.feifel@physics.gu.se

## Introduction

$SF_6$ is a highly symmetric molecule in its neutral ground state, exhibiting a regular octahedral structure that belongs to the $O_h$ point group. This geometry is very stable, giving $SF_6$ a high dielectric strength while at the same time being non-toxic and non-flammable. These properties make it a prime choice for use in the electrical power industry, which consumes roughly 80 % of all $SF_6$.(Palmer 1996; MALLER and NAIDU 1981) Despite its stability, fragmentation of the molecule upon ionization leads to $SF_n$ species ($n < 6$) that are toxic and corrosive.(MALLER and NAIDU 1981) Related to this, $SF_6$ is also classified as a greenhouse gas, further motivating detailed studies of its structure and dynamics.(Palmer 1996)

Given its broad relevance, it is important to understand the physical and chemical properties of $SF_6$ in different charge and excitation states, in which its electronic structure plays a central role. For the neutral ground state, comprising 70 electrons, the electron configuration can be written as

$[1a_{1g}^2(S1s^2)\ 2a_{1g}^2 1e_g^4 1t_{1u}^6(F1s^{12})\ 3a_{1g}^2(S2s^2)\ 2t_{1u}^6(S2p^6)]\ \{4a_{1g}^2 3t_{1u}^6 2e_g^4(F2s^{12})\}\ 5a_{1g}^2 4t_{1u}^6 1t_{2g}^6\ 3e_g^4\ 5t_{1u}^6\ 1t_{2u}^6\ 1t_{1g}^6\ (F2p^{36})$.

Here, the square brackets indicate the core orbitals, the curly brackets the inner-valence orbitals, and the remaining part reflects the outer-valence orbitals.

Ionization of $SF_6$ has been the subject of numerous studies, with conventional electron spectra reported for both the valence and inner-shell regions. A comprehensive summary of the literature up to 2004 is provided in Ref.(Feifel et al. 2005), which also presented the first complete double-valence-ionization electron spectra of $SF_6$. These spectra were found to exhibit essentially no fine structure despite the comparatively high electron-energy resolution used for the measurements. This was attributed to the congestion of a high number of dicationic states.

The fragmentation dynamics following single ionization have been investigated extensively,(Bull et al. 2017) and, in this context, double electron detachment of $SF_6^-$ is also relevant.(Kandhasamy et al. 2015) Most recently, these efforts were complemented by a multi-electron–ion coincidence study of $SF_6^{2+}$ dissociation, supported by molecular dynamics simulations and $M_3C$ statistical calculations.(Olsson, Daver Ideböhn, et al. 2025) Apart from that, recent $SF_6$ works investigated structural changes of this system upon S1s inner-shell excitation using either sulphur $L$-shell emission spectroscopy or cold target recoil-ion momentum spectroscopy (COLTRIMS) for detection.(McGinnis et al. 2026; Travnikova et al. 2025)

In the present work, we extend the understanding of the electronic structure of $SF_6$ by focusing on its core–valence double ionization, measured above the S2$p$, F1$s$, and S1$s$ ionization thresholds, respectively. Earlier studies of core–valence double ionization of molecular systems using electron–electron coincidence spectroscopy (Andersson et al. 2008, 2010; Olsson, Cornetta, et al. 2025) demonstrated that the resulting spectra can closely resemble the corresponding single-ionization valence photoelectron spectra. This similarity indicates that the overall Coulomb interaction can dominate over more specific hole–hole coupling mechanisms, even when the core orbitals are split by spin–orbit interaction.(Andersson et al. 2010) From a theoretical point of view, such valence-like behavior is expected when the core orbital has limited spatial overlap with the valence orbitals.

Deviations from valence-like behaviour can arise when symmetry breaking becomes important, as first demonstrated for $SO_2$, where a (pseudo-)Jahn–Teller effect was invoked,(Niskanen et al. 2012) and more recently for the highly symmetric allene molecule(Ideböhn et al. 2023). Additionally, strong site-selective effects may occur, as reported very recently for the case of carbon suboxide. This aspect becomes particularly interesting for molecules with symmetrically arranged ligands such as $SF_6$, where ionization from qualitatively different atomic sites can be expected to yield distinct spectral signatures - an aspect that forms one of the central themes of the present study.

# Experiments

The experiments were carried out using time-of-flight photoelectron–photoelectron coincidence (TOF-PEPECO) spectroscopy both at the synchrotron radiation facility PETRA III at DESY in Hamburg and in our laboratory in Gothenburg. In the home laboratory, a 5.6 m long time-of-flight magnetic-bottle spectrometer was used to record single-ionization electron spectra using 21 and 41 eV photons from a pulsed helium gas-discharge lamp operated at a repetition rate of 4.4 kHz. The method has been described in detail in previous works, including the original report by one of us and co-workers.(Eland et al. 2003) In brief, a conical pole piece mounted on a permanent magnetic rod generates a strong magnetic field of approximately 1 T at its tip. This field couples into a homogeneous guiding field of a few mT created by a solenoid along the flight tube. The resulting configuration is known to collect essentially all electrons produced in the ionization process, including those with kinetic energies of several hundred electronvolts, as relevant for the synchrotron-based measurements. The 5.6 m TOF-PEPECO instrument provides a collection–detection efficiency of about 40 % and a nominal resolving power of $E/\Delta E = 120$. The photon energy from the helium discharge lamp is selected by an ellipsoidal grating with a groove density of 1100 lines/mm.

Measurements at photon energies above 41 eV were performed at beamline P04 of PETRA III using our mobile 2.2 m TOF-PEPECO spectrometer. This setup is essentially identical to the one used in Gothenburg, except for the light source and its total flight-tube length. A mechanical chopper system synchronized to the radiofrequency of the PETRA III storage ring increased the inter-pulse spacing from 192 ns in 40-bunch mode to approximately 10 $\mu$s, ensuring that electrons with long flight times are detected before the next light pulse arrives(Viefhaus et al. 2013). The ionization rate was kept below 2 % to ensure that electrons originating from the same ionization event form the basis of the correlation features. The 2.2 m instrument has a collection–detection efficiency of about 50 % and a nominal resolving power of $E/\Delta E = 50$.

The gaseous sample was obtained commercially from Air Liquide, with a stated purity of 99.9 %. It was introduced into the spectrometers through hollow stainless-steel needles with an inner diameter of 1 mm, producing effusive molecular beams.

# Theory

## Theoretical model for the core-valence DI intensities

The single-photon ($\gamma$) double-ionization (DI) reaction

$$\gamma + \mathrm{M} \;\rightarrow\; \mathrm{M}_f^{2+} + e^-(\boldsymbol{k}_1) + e^-(\boldsymbol{k}_2)$$

is treated within the electric-dipole approximation and first-order perturbation theory in the photon–molecule coupling. Energy conservation,

$$\omega = \mathrm{DIP}_f + \varepsilon_1 + \varepsilon_2, \qquad \mathrm{DIP}_f = E_f^{(N-2)} - E_0^{(N)},$$

defines the double-ionization potential $\mathrm{DIP}_f$ of the dication state $f$ and the excess energy

$$E_{\text{excess}} \equiv \omega - \mathrm{DIP}_f = \varepsilon_1 + \varepsilon_2 \geq 0$$

shared between the two photoelectrons with kinetic energies $\varepsilon_s = k_s^2/2$ and asymptotic momenta $\boldsymbol{k}_s$ ($s = 1{,}2$). The double-photoionization channel $f$ is open only when $\omega$ is larger than $\mathrm{DIP}_f$, which here is calculated within the ΔSCF (Self-Consistent Field) approach.

Two conceptually distinct physical mechanisms are responsible for the DI process, and they operate with different relative weights depending on the photon energy and the system. In the *shake-off* mechanism, the photon ejects one electron via a dipole transition. Because this process is sudden compared to the relaxation timescale of the remaining electrons, the sudden change in the mean field projects the $(N-1)$-electron residual wavefunction onto the eigenstates of the $(N-1)$-electron Hamiltonian. Formally, in the strict independent-electron picture, this probability is exactly zero. It becomes nonzero when the orbitals change upon ionization, which is itself a measure of the breakdown of the frozen-orbital approximation. The shake-off mechanism dominates at high photon energies, where the primary electron departs rapidly and the sudden approximation is justified. Despite the necessity of orbital relaxation for the shake-off probabilities to be nonzero, this is not true for the plane-wave approximation, due to the nonzero overlap between plane waves and the atomic orbitals. Deviations from the frozen-orbital limit are discussed in the Results section.

Alternatively, in the *knock-out* mechanism, the photon ejects one electron, which then scatters off a second bound electron via Coulomb repulsion. This mechanism requires an explicit treatment of the electron–electron interaction between the primary photoelectron and the bound electrons, and is therefore absent in any independent-electron or mean-field description of the continuum or in the plane-wave approximation. Although the knock-out mechanism tends to have a significant contribution at intermediate photon energies and to produce characteristic energy-sharing signatures, the present work focuses on the shake-off mechanism alone. Since the relative weight of the knock-out contribution is expected to be larger for smaller excess photon energies above threshold, we should expect that, for a fixed photon energy, the shake-off approximation will be less accurate for CV states with the highest double-ionization energies.

In the following, we present the approximations of the model in detail for (i) the S1$s$ and F1$s$ cases and (ii) S2$p$ case. General aspects of the model are discussed alongside the specifics of each section.

## The S1$s$ and F1$s$ edges

The electronic structure of the neutral ground state $|\Psi_0\rangle$ of the closed-shell molecule M ($N = 2n$ electrons) is described at the restricted Hartree–Fock (RHF) level, built from $n$ doubly occupied spatial molecular orbitals $\{\phi_i\}_{i=1}^{n}$ obtained from the Fock operator of the neutral system, $\hat{F}^{(N)}\phi_i = \varepsilon_i\,\phi_i$. The CV dication $\mathrm{M}_f^{2+}$ produced after DI carries one hole in a core orbital $\varphi_i$ and one hole in a valence orbital $\varphi_j$; its electronic structure is described at the open-shell restricted Hartree–Fock (OSRHF) level with a separate set of orbitals $\{\varphi_k\}_{k=1}^{n-1}$, relaxed in the dication mean field, $\hat{F}^{(N-2)}\varphi_k = \eta_k\,\varphi_k$. The two ejected electrons occupy continuum orbitals $\psi_{\boldsymbol{k}_1}$ and $\psi_{\boldsymbol{k}_2}$. Because $\hat{F}^{(N)}$ and $\hat{F}^{(N-2)}$ describe distinct mean fields, the two orbital sets are not mutually orthogonal, $\langle\varphi_k|\phi_i\rangle = S_{ki} \neq \delta_{ki}$, and this non-orthogonality is the very mechanism through which orbital relaxation contributes to the DI amplitude.

For the S1$s$ and F1$s$ edges, the $LS$ coupling scheme provides a convenient basis. It means that, for a closed-shell singlet target, total-spin conservation ($\Delta S = 0$, $\Delta M_S = 0$) selects, for each hole pair $(i, j)$, two distinct CV final states: a singlet dication ($S_{\text{dic}} = 0$) coupled to a singlet electron pair ($S_{\text{pair}} = 0$), and a triplet dication ($S_{\text{dic}} = 1$) coupled to a triplet pair ($S_{\text{pair}} = 1$). The properly antisymmetrized $N$-electron final states are

$$\left|\Psi_f^{S_{\text{dic}}=0}\right\rangle = \frac{1}{2}\Big(\left|\Phi^{(i\alpha,j\beta)}\,\psi^{\alpha}_{\mathbf{k}_1}\psi^{\beta}_{\mathbf{k}_2}\right\rangle - \left|\Phi^{(i\alpha,j\beta)}\,\psi^{\beta}_{\mathbf{k}_1}\psi^{\alpha}_{\mathbf{k}_2}\right\rangle - \left|\Phi^{(i\beta,j\alpha)}\,\psi^{\alpha}_{\mathbf{k}_1}\psi^{\beta}_{\mathbf{k}_2}\right\rangle + \left|\Phi^{(i\beta,j\alpha)}\,\psi^{\beta}_{\mathbf{k}_1}\psi^{\alpha}_{\mathbf{k}_2}\right\rangle\Big), \tag{4}$$

$$\begin{aligned}\left|\Psi_f^{S_{\text{dic}}=1}\right\rangle = {} & \frac{1}{\sqrt{3}}\left|\Phi^{(i\alpha,j\alpha)}\,\psi^{\alpha}_{\mathbf{k}_1}\psi^{\alpha}_{\mathbf{k}_2}\right\rangle + \frac{1}{\sqrt{3}}\left|\Phi^{(i\beta,j\beta)}\,\psi^{\beta}_{\mathbf{k}_1}\psi^{\beta}_{\mathbf{k}_2}\right\rangle \\ & - \frac{1}{2\sqrt{3}}\Big(\left|\Phi^{(i\alpha,j\beta)}\,\psi^{\alpha}_{\mathbf{k}_1}\psi^{\beta}_{\mathbf{k}_2}\right\rangle + \left|\Phi^{(i\beta,j\alpha)}\,\psi^{\alpha}_{\mathbf{k}_1}\psi^{\beta}_{\mathbf{k}_2}\right\rangle + \left|\Phi^{(i\alpha,j\beta)}\,\psi^{\beta}_{\mathbf{k}_1}\psi^{\alpha}_{\mathbf{k}_2}\right\rangle + \left|\Phi^{(i\beta,j\alpha)}\,\psi^{\beta}_{\mathbf{k}_1}\psi^{\alpha}_{\mathbf{k}_2}\right\rangle\Big),\end{aligned} \tag{5}$$

where $\left|\Phi^{(i\sigma,j\sigma')}\,\psi^{\sigma_1}_{\boldsymbol{k}_1}\psi^{\sigma_2}_{\boldsymbol{k}_2}\right\rangle$ denotes an antisymmetric $N$-electron Slater determinant built from the $(N-2)$-electron dication configuration — with holes of specified spin in the $\varphi_i$ and $\varphi_j$ MOs — coupled by the pair of continuum spin-orbitals. The singlet and triplet final states have distinct dication energies (and hence distinct to double-ionization potentials).

The transition amplitude $T_f = \langle\Psi_f|\hat{\boldsymbol{\epsilon}}\cdot\hat{\boldsymbol{r}}|\Psi_0\rangle$ for each CV state of a given spin multiplicity of the dication, given by either Eq. [eq:singlet_state] or [eq:triplet_state], is a sum of transition amplitudes of individual *channels* - or *components* - of the form $T_f^{(i\sigma,j\sigma'),\sigma_1\sigma_2} = \langle\Phi^{(i\sigma,j\sigma')}\,\psi^{\sigma_1}_{\boldsymbol{k}_1}\psi^{\sigma_2}_{\boldsymbol{k}_2}|\hat{\boldsymbol{\epsilon}}\cdot\hat{\boldsymbol{r}}|\Psi_0\rangle$. The transition amplitude of a given component is evaluated through the generalized Slater–Condon rules for non-orthogonal determinants, which naturally split $T_f$ into a *direct* contribution - the dipole operator promotes a bound electron to the continuum while the second electron is emitted by shake-off - and an *indirect* contribution - the dipole operator drives a bound-to-bound transition within the dication, while the sudden change of mean field is responsible for the two electron ejections. In other words, the amplitude can be written as $T_f = T_f^{\text{dir}} \pm T_f^{\text{ind}}$, which makes the differential cross section proportional to $|T_f^{\text{dir}}|^2 + |T_f^{\text{ind}}|^2 \pm 2\text{Re}[T_f^{\text{dir}}T_f^{\text{ind}*}]$, where the latter term corresponds to the interference between the direct and indirect amplitudes.

Inspired by the Gelius model for single ionization(Gelius 1974), in which the molecular photoemission intensity is expressed as a weighted sum of atomic sub-shell cross sections under a one-centre LCAO projection, we here (i) retain only the direct contribution and (ii) neglect off-centre terms. First, the direct one-centre terms have the advantage of admitting a clean reduction to tabulated atomic quantities, and the resulting working expressions for the DI cross sections share the same one-centre spirit as the original Gelius single-ionization model. Secondly, further numerical inspections show that the ratio between the intensity from the off-centre direct plus all indirect terms and the intensity from the direct one-center terms is less than $10^{-4}$ for all given CV states, which justifies this part of the approximation. However, the interference between the direct and indirect terms, although neglected here, is not in principle negligible, and this ratio relative to the direct one-center intensity can reach $10^{-2}$ to $10^{-1}$, being larger for the lowest-lying triplet CV states.

We now present how the intensity is computed within our approximation. Two classes of objects central to the model are introduced here. The first is the one-electron Dyson orbital associated with hole $i$, projected onto the AO basis $\{\chi_\mu\}$,

$$|d_i\rangle \;=\; \sum_q \mathfrak{d}_i^q \;|\phi_q\rangle \;=\; \sum_\mu d_i^\mu \;|\chi_\mu\rangle, \qquad d_i^\mu \;=\; \sum_q \mathfrak{d}_i^q \; c_{\mu q},$$

where $c_{\mu q}$ are the MO coefficients of the ground state orbitals, the AO weights $d_i^\mu$ encode the orbital-relaxation response of the system upon removal of an electron from $\varphi_i$, and whose squared norm $P_i = \sum_\mu (\, d_i^\mu)^2 \leq 1$ is the usual spectroscopic factor of hole $i$. The second is the two-electron Dyson amplitude. It is important to note that the two-electron contribution survives in the $\alpha\alpha/\beta\beta$ channel of the triplet, in which both photoelectrons are emitted into the continuum with the same spin. This channel is governed by the two-electron Dyson amplitudes $\mathfrak{D}_{ij}^{pq}$, which encode the joint response of the system to the simultaneous removal of two electrons from orbitals $\varphi_i$ and $\varphi_j$, and which project onto pairs of AOs as

$$D_{ij}^{\mu\nu} \;=\; \sum_{p<q} \mathfrak{D}_{ij}^{pq} \; c_{\mu p}\, c_{\nu q}.$$

Within the one-centre approximation, the angle-averaged direct amplitude squared reduces to combinations of two one-electron building blocks,

$$\mathcal{D}_i = \mathcal{D}_i(\varepsilon_1) \simeq \sum_\mu |\, d_i^\mu|^2 \,\frac{\sigma_\mu^{\mathrm{AO}}(\varepsilon_1 + I_\mu)}{(\varepsilon_1 + I_\mu)},$$
$$\mathcal{S}_j = \mathcal{S}_j(\varepsilon_2) \simeq \sum_\nu |\, d_j^\nu|^2 \, k_2 \, P_\nu^{\mathrm{shake}}(k_2),$$

where $\sigma_\mu^{\mathrm{AO}}$ is the atomic sub-shell photoionisation cross section taken from Yeh and Lindau(Yeh and Lindau 1985), $I_\mu$ is the corresponding atomic sub-shell threshold, $P_\nu^{\mathrm{shake}}$ is the differential shake-off probability of AO $\chi_\nu$, and $k_s = \sqrt{2\varepsilon_s}$, and one two-electron block,

$$\mathcal{G}_{ij} = \mathcal{G}_{ij}(\varepsilon_1, \varepsilon_2) \simeq \; |\det(\boldsymbol{S})|^2 \sum_{\mu\nu} |\, D_{ij}^{\mu\nu}|^2 \,\frac{\sigma_\mu^{\mathrm{AO}}(\varepsilon_1 + I_\mu)}{(\varepsilon_1 + I_\mu)} \; k_2 \, P_\nu^{\mathrm{shake}}(k_2),$$

where $\boldsymbol{S}$ is the overlap matrix of the dication bound-orbital set against the neutral-orbital set. The per-AO effective photon energy $\varepsilon_1 + I_\mu$ entering $\mathcal{D}_i$ ensures that each atomic cross section is evaluated near its own threshold. The two quantities have a clear physical interpretation: $\mathcal{D}_i$ is the atomic-resolved photoionization strength of hole $i$, weighted by the AO content of its Dyson orbital, while $\mathcal{S}_j$ is the corresponding shake-off yield. The product $\mathcal{D}_i\mathcal{S}_j$ then describes photoionization of hole $i$ combined with shake-off of hole $j$, which is the dominant mechanism for CV states. On the other hand, $\mathcal{G}_{ij}$ is the two-electron analogue of $\mathcal{D}_i\mathcal{S}_j$, where both continuum electrons now appear simultaneously through their overlap with the neutral, while the dipole acts within the AO pair $(\mu, \nu)$.

The differential shake-off probability $P_\nu^{\text{shake}}(k)$ entering $\mathcal{S}_i$ and $\mathcal{G}_{ij}$ is the probability density for an electron originally in AO $\chi_\nu$ to be projected, by the sudden change of mean field upon ionization, onto a continuum state of momentum $\boldsymbol{k}$. In the plane-wave approximation, and after averaging over the direction of $\boldsymbol{k}$,

$$P_\nu^{\text{shake}}(k) \;=\; \frac{1}{(2\pi)^3}\,\langle|\tilde{\chi}_\nu(\boldsymbol{k})|^2\rangle_{\Omega_{\boldsymbol{k}}}, \qquad \tilde{\chi}_\nu(\boldsymbol{k}) = \int e^{-i\boldsymbol{k}\cdot\boldsymbol{r}}\,\chi_\nu(\boldsymbol{r})\,d^3r,$$

where $\langle . \rangle_{\Omega_{\boldsymbol{k}}}$ denotes the average over the orientation of $\boldsymbol{k}$. $P_\nu^{\text{shake}}$ is a purely atomic quantity, computed once for each AO of the basis set from its analytic Fourier transform, and carries the shake-off mechanism throughout the calculation pipeline.

After angular integration over $\Omega_{\boldsymbol{k}_1} and \Omega_{\boldsymbol{k}_2}$, and averaging over the photon polarization, the singly differential DI cross section, differentiated by $\varepsilon_1$, for the final state $f$ reads

$$\frac{d\sigma_f}{d\varepsilon_1} \;=\; \frac{4\pi^2\omega}{3c}\,k_1k_2\,A_f(\varepsilon_1,\,E_{\text{excess}} - \varepsilon_1),$$

where $\omega$ is the photon frequency. The full angle-averaged squared amplitudes for the singlet and triplet final states include, in addition to $\mathcal{D}_i\mathcal{S}_j$ and $\mathcal{G}_{ij}$, a set of cross-Dyson interference terms that couple the two holes within a single channel and across the $\alpha\beta$ and $\alpha\alpha/\beta\beta$ channels. For CV states, these terms are suppressed by the near-orthogonality of the core and valence Dyson orbitals in the AO basis and can be safely neglected at the level of accuracy targeted here. The squared amplitudes then reduce to

$$\begin{aligned} k_1k_2A_f^{(S_{\text{dic}}=0)} &\;\simeq\; 2\big(\mathcal{D}_i\mathcal{S}_j + \mathcal{D}_j\mathcal{S}_i\big) = 2\mathcal{F}_{ij}, \\ k_1k_2A_f^{(S_{\text{dic}}=1)} &\;\simeq\; \frac{1}{3}\big[\,2\big(\mathcal{D}_i\mathcal{S}_j + \mathcal{D}_j\mathcal{S}_i\big) + 4\mathcal{G}_{ij}\,\big] = \frac{2}{3}\big(\mathcal{F}_{ij} + 2\mathcal{G}_{ij}\big), \end{aligned}$$

where $\mathcal{F}_{ij} = \mathcal{D}_i\mathcal{S}_j + \mathcal{D}_j\mathcal{S}_i$ for brevity. In this form, the leading singlet and triplet intensities are driven by the same one-electron object $\mathcal{D}_i\mathcal{S}_j + \mathcal{D}_j\mathcal{S}_i$, while the triplet acquires an extra two-electron contribution $\mathcal{G}_{ij}$ that accounts for the $\alpha\alpha$ and $\beta\beta$ components. The two final states give similar intensities for CV configurations, and the underlying balance can be read off the structure of Eqs. [eq:A_singlet] and [eq:A_triplet]. The singlet intensity is carried entirely by the $\alpha\beta$ direct channel, $2(\mathcal{D}_i\mathcal{S}_j + \mathcal{D}_j\mathcal{S}_i)$. The triplet receives the same $\alpha\beta$ contribution, down-weighted by the Clebsch–Gordan factor $1/3$ of the unique $S = 0, M_S = 0$ spin coupling, and acquires in compensation the $\alpha\alpha/\beta\beta$ direct two-continuum channels $\mathcal{G}_{ij}$, which is closed for the singlet. The two contributions are governed by the same one-electron Dyson amplitudes built on the same pair of holes $(i, j)$ and, for CV states, are of comparable magnitude. The photon-ionization and shake-off factors that drive $\mathcal{D}_j\mathcal{S}_i$ in the $\alpha\beta$ channel reappear, with analogous ingredients, inside $\mathcal{G}_{ij}$ in the $\alpha\alpha/\beta\beta$ channel. Interestingly, the Clebsch–Gordan penalty on the $\alpha\beta$ direct channel and the gain from the $\alpha\alpha/\beta\beta$ channel therefore approximately cancel, and the model predicts a singlet-to-triplet branching ratio close to unity at each $(i, j)$.

The total cross section for a given CV state follows from the integral over the energy-sharing variable,

$$\sigma_f^{\mathrm{DI}}(\omega) = \frac{4\pi^2\omega}{3c}\int_0^{E_{\mathrm{excess}}} k_1 k_2\, A_f(\varepsilon_1, E_{\mathrm{excess}} - \varepsilon_1)\, d\varepsilon_1,$$

and the simulated CV spectrum at photon energy $\omega$ is obtained by distributing the $\sigma_f^{\mathrm{DI}}$ over their respective dication energies and convolving with a phenomenological line profile that accounts for vibrational broadening, the lifetime of the CV states, and the experimental resolution, here modelled with a Voigt profile. The resulting simulated spectra are compared with the measured ones in Figs. 2 **(a)** and 3 **(a)**, with each integral listed in Tables 1 and 3. The agreement validates the OSRHF treatment of the CV final states and the direct-channel intensity model adopted here for the outer-valence region. The deviations observed for the inner-valence part of the spectrum, especially for the F1$s$ case, are consistent with the onset of MO breakdown mentioned above, which is beyond the description of the dication adopted here.

## The S2$p$ edge

Unlike the F1$s$ and S1$s$ edges considered above, where the spin–orbit (SO) coupling within the core hole is zero at first order and the dicationic two-hole configuration is naturally described in the $LS$-coupling scheme, at the S2$p$ edge this picture needs to be modified. The SO splitting of the S2$p$ orbital is $\Delta_{\mathrm{SO}} \approx 1.2$ eV, which is comparable to or larger than the exchange splittings of the CV states. SO coupling of the valence hole and of the two ejected photoelectrons, on the other hand, remains negligible. We therefore propose an intermediate-coupling regime, in which the core hole is coupled internally by $\hat{H}_{\mathrm{SO}}$ while the rest of the system retains its $LS$-coupled structure. SO coupling on the core hole couples its orbital and spin angular momenta, so the good single-hole quantum number for the core becomes $j_C = 1/2, 3/2$, each with $2j_C + 1$ sublevels. Coupling $j_C$ with the valence-hole spin $1/2$ then yields the dication total angular momentum, resulting in four allowed dication states per hole pair of CV MOs are therefore $(j_C, J_{\mathrm{dic}}) \in \{(3/2,\ 2),\ (3/2,\ 1),\ (1/2,\ 1),\ (1/2,\ 0)\}$. We emphasise that $J_{\mathrm{dic}}$ here is not the actual total angular momentum of the dication, as the valence MO character is treated as a spectator under the partial coupling. Rather, it can be understood as the quantum number that captures how the atomic SO splitting of the core hole combines with the valence-hole spin.

The selection rules acting on the full $N$-electron final state are unchanged from the $LS$-coupled cases, i.e., the final state must have total electronic spin $S = 0$. What changes at the S2$p$ edge is the way these constraints are adapted. The value of $J_{\mathrm{dic}}$ can be 0, 1, or 2, with all four $(j_C, J_{\mathrm{dic}})$ dication states therefore being accessible. As obtained from the 6-$j$ symbols, the two states $(j_C, J_{\rm dic}) = (3/2{,}2)$ and $(1/2{,}0)$ are specific Clebsch–Gordan projections within the $|S_{\rm dic} = 1\rangle$ spin manifold, built from the three $M_L^{\rm dic}$ components of the S $2p$ core hole. Because the $|S_{\rm dic} = 0\rangle$ component vanishes for both such states, their angle-averaged squared amplitudes are obtained from the $LS$-triplet amplitude $A_f^{(S=1)}$ alone. The explicit $N$-electron Slater-determinant expansions of all $(j_C, J_{\rm dic})$ states are presented in the Supplementary Material.

Given these wave functions, we can work out the angle-averaged expressions for the squared transition amplitude. First, we note that, because the $\alpha\beta$ and $\alpha\alpha/\beta\beta$ direct channels involve different spin configurations of the two continuum electrons, the singlet–triplet interference vanishes identically in the simplified one-center approximation. Next, we note that, upon neglecting the cross terms, the squared amplitude for each $(j_C, J_{\rm dic})$ state reduces to a linear combination of $A_f^{(S_{\rm dic}=0)}$ and $A_f^{(S_{\rm dic}=1)}$. The resulting angle-integrated transition amplitudes for the two unresolved $j_C$ peaks are therefore

$$
\begin{aligned}
k_1 k_2 A_f^{(j_C=3/2)} &= \frac{8}{3}\,\mathcal{F}_{ij} + \frac{4}{3}\,\mathcal{G}_{ij}, \\
k_1 k_2 A_f^{(j_C=1/2)} &= \frac{4}{3}\,\mathcal{F}_{ij} + \frac{2}{3}\,\mathcal{G}_{ij}.
\end{aligned}
$$

Both amplitudes are proportional to the same combination $2\mathcal{F}_{ij} + \mathcal{G}_{ij}$, with the $j_C = 3/2$ peak twice as intense as the $j_C = 1/2$ peak. In terms of the $LS$ amplitudes, Eqs. [eq:Aj32_supp]–[eq:Aj12_supp] read

$$
\begin{aligned}
A_f^{(j_C=3/2)} &= A_f^{(S_{\rm dic}=0)} + A_f^{(S_{\rm dic}=1)}, \\
A_f^{(j_C=1/2)} &= \frac{1}{2}\Big[A_f^{(S_{\rm dic}=0)} + A_f^{(S_{\rm dic}=1)}\Big].
\end{aligned}
$$

This is the form used in the code to combine the two pipeline outputs $A_f^{(S=0)}$ and $A_f^{(S=1)}$ into the two $j_C$ peaks once SO coupling on the core hole is switched on.

## The frozen-orbital limit

Alongside the ΔSCF result, we also report the frozen-orbital-limit reference spectrum, whose role is essentially diagnostic. In this limit, the dication's occupied MOs coincide with those of the closed-shell neutral ground state, obtained from the RHF reference calculation of the parent molecule. The $N-2$ electrons of the dication are simply placed into the neutral occupied MOs, with two specific orbitals left singly occupied - the core $i$ and valence $j$ - and no ΔSCF re-optimization is performed. When the dication MOs set coincides with the neutral one, the one- and two-electron Dyson objects between the neutral ground state and each dicationic CV state in Eqs. [eq:dyson_orbital] and [eq:two_electron_dyson] reduce to $\mathfrak{d}_i^q \to \delta_{q,i}$ and $\mathfrak{D}_{ij}^{pq} \to \big(\delta_{p,i}\,\delta_{q,j} - \delta_{p,j}\,\delta_{q,i}\big)$.

## Computational details

Earlier core-valence spectra have been analyzed by a variety of models(Ideböhn et al. 2023; Olsson, Cornetta, et al. 2025; Zagorodskikh et al. 2016), including configuration interaction, complete and restricted active-space perturbation theory, and OSRHF. As $SF_6$ is “electron-rich”, with 36 valence electrons, explicitly correlated methods will, despite the high symmetry, be restricted towards the lower eigenstates of the core-valence spectrum. As many of these constitute satellite states with small cross sections, effectively only a limited part of the spectrum will be covered for a molecule like $SF_6$. OSRHF, as described by the smallest possible expansion fulfilling spatial and spin symmetry, is, on the other hand, computationally effective and can cover the full CV spectrum. It thus excludes electron-correlation effects which become progressively more important for deeper parts of the CV spectrum and may lead to overestimated calculated band energies in these regions. However, as it implements a (spin- coupled) independent-particle model with the possibility of reaching the full basis-set limit, it provides a rigorous framework for analyzing CV spectra.

All electronic structures of the CV states at the OSRHF level of theory were obtained using the Restricted Active-Space Self-Consistent-Field (RASSCF) routine of the OpenMOLCAS package(Fdez. Galván et al. 2019), together with the cc-pVDZ basis set. The choice of the basis set was based on the character of the atomic orbitals, since they are to be associated with the $\sigma_\mu^{\rm AO}$ quantities of Eqs. [eq:Di] and [eq:Gij]. The energy dependences of the atomic cross sections were obtained using a cubic spline interpolation of the tabulated data(Yeh and Lindau 1985). Although $SF_6$ belongs to the $O_h$ point group, no symmetry constraints were imposed throughout the calculations. Particularly for the F1$s$ CV spectrum, localized MOs based on the Pipek-Mezey(Pipek and Mezey 1989) procedure were used as input orbitals for the optimization of the CV states, and the $O_h$ labels assigned to the orbitals were obtained via a post-processing analysis. The SO coupling and the SO splitting of the S2$p$ level for the S2$p$ CV spectrum were calculated using the Restricted Active-Space State Interaction (RASSI) program, as implemented in OpenMOLCAS.

# Results and Discussion

Figure 1 shows double-ionization electron-pair spectra taken above the S2$p$ (180.4 eV & 181.7 eV), S1$s$ (2491 eV), and F1$s$ (694.6 eV) core-ionization thresholds(Siegbahn et al. 1969), respectively, at the photon energies stated in the figure. For comparison, the well-known, conventional single-ionization valence photoelectron spectra obtained at the photon energies of 21 and 41 eV, respectively, are also included (which agree very well with previously published results).(Holland et al. 1995) The resolution in the spectrum taken at 21 eV is about 0.05 eV, which is sufficient to resolve vibrational fine structure in the D $^2T_{2g}$ state around 20 eV binding energy, where the vibrational spacing is known to be about 0.07 eV. (Holland et al. 1995) In the 41 eV spectrum shown in the upper panel, the resolution for the X-state is about 0.2 eV, and the spectrum includes several more electronic states compared to the 21 eV spectrum (lowest panel). The vertical ionization energies in the 41 eV spectrum are found at 15.7, 17, 18.4, 19.8, 22.5, and 26.8 eV, all listed in Table 11 together with the other visible peaks in the figure. As seen in Fig. 1, the spectrum stretches over about 15 – 40 eV, and its six distinct bands can be characterized by ionization from seven outer orbitals of F2$p$, S3$s$, and S3$p$ character: 5a$_{1g}$, 4t$_{1u}$, 1t$_{2g}$, 3e$_{g}$, 5t$_{1u}$, 1t$_{2u}$, and 1t$_{1g}$ (Holland et al. 1995). Also noted in Ref. (Holland et al. 1995) are three inner valence MOs of F2$s$ character, 4a$_{1g}$, 3t$_{1u}$, and 2e$_{g}$, which are separated from the outer valence ionizations by a large gap (12 eV), and appear at around 40 eV. The inner-valence bands are broad and faint compared to the outer-valence peaks, as they most probably are subject to strong configurational splittings with associated MO breakdown, while the outer-MO bands are relatively sharp and strong and are largely well described by an independent-particle model (with the exception noted below), as implemented here by OSRHF.

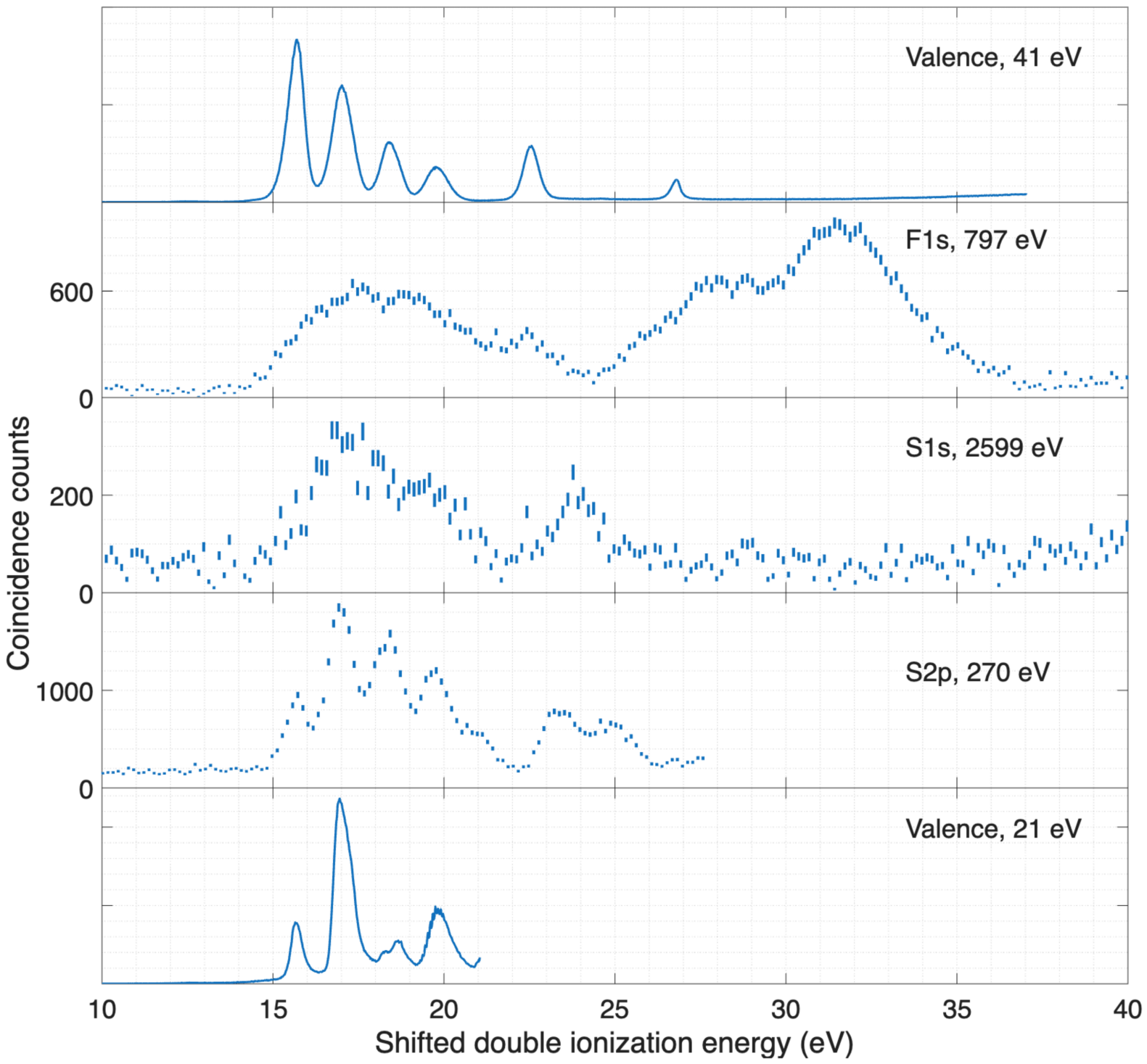

*Core-valence double-ionization electron-pair spectra of $SF_6$ obtained at photon energies well above the core S2p , F1s, and S1s energies at 270, 797, and 2599 eV, respectively, in comparison to the well-known single-ionization valence electron spectra obtained at 21 eV (lowest panel), and 41 eV (highest panel). The error bars in the plots reflect the statistical uncertainty, except for the valence electron spectra, which are plotted without error bars owing to the high number of counts obtained.*

As can be seen in the second panel from the bottom, the S2$p$ core-valence spectrum obtained at a photon energy of 270 eV is very sharp and closely resembles the conventional valence photoelectron spectrum recorded at 21 eV. Similarly, as presented in the middle panel of this figure, the core-valence double-ionization spectrum taken above the S 1s edge using a photon energy of 2599 eV also resembles the single-ionization valence photoelectron spectrum obtained at 21 eV in terms of the energetic positions of the features and their intensity distribution, when taking into account the difference in resolution, which is about 2 eV for the first group of states.

All bands in the experimental spectra have been assigned according to the calculations presented in Figs. 2 **(b)**, 3 **(a)** and 3 **(b)**, and summarized in Table 11. The individual assignments for the calculated peaks in Figs. 2 **(b)**, 3 **(a)** and 3 **(b)** are found in Tables 3,1, and 2, respectively. Table 3 is representative of the data presented in Fig. 2 **(a)**, using localized MOs where the $O_h$ symmetry breaks, and the energies are therefore assigned only triplets or singlets. Calculations using delocalized MOs, preserving the molecular symmetry, have also been performed for each F1$s$ ionization site using both frozen-orbital sum and the optimized-MOs methods discussed above, and are plotted in Fig. 2 **(b)** and in Figs. S4 and S5 of the Supplementary Materials. These calculations form the basis of the electronic-state assignments of the F1$s$ CV spectrum presented in Tables 3 and 11.

In the fourth panel of Fig. 1, the core-valence double-ionization spectrum recorded above the F1$s$ edge is shown. This spectrum was taken at 797 eV, which is about 100 eV above the F1$s$ edge(Siegbahn et al. 1969). It appears rather different from the other two core-valence spectra. The spectrum is shifted by slightly more than the F1$s$ binding energy to enable a direct comparison with the single-ionization valence photoelectron spectrum and with the other core-valence spectra. The resolving power at this energy is about 1.6 eV, which explains the limited resolution compared to the single-ionization spectrum.

As in the single-ionization spectrum and the other two core-valence spectra, the first fairly broad structure in the F1$s$ spectrum likely contains contributions from the first six outer-valence orbitals. However, because of the lower resolution, the individual peaks are less well resolved, although this first band clearly covers the energy range of nearly all of the outer-valence cationic states. In strong contrast - particularly to the single-ionization valence spectrum obtained at 41 eV - additional spectral features appear in this core-valence spectrum at ionization energies of 30.4 eV and higher. The strong intensity in the 25 eV and higher ionization-energy region may be due to contributions involving the $5a_{1g}^2$ valence orbital.

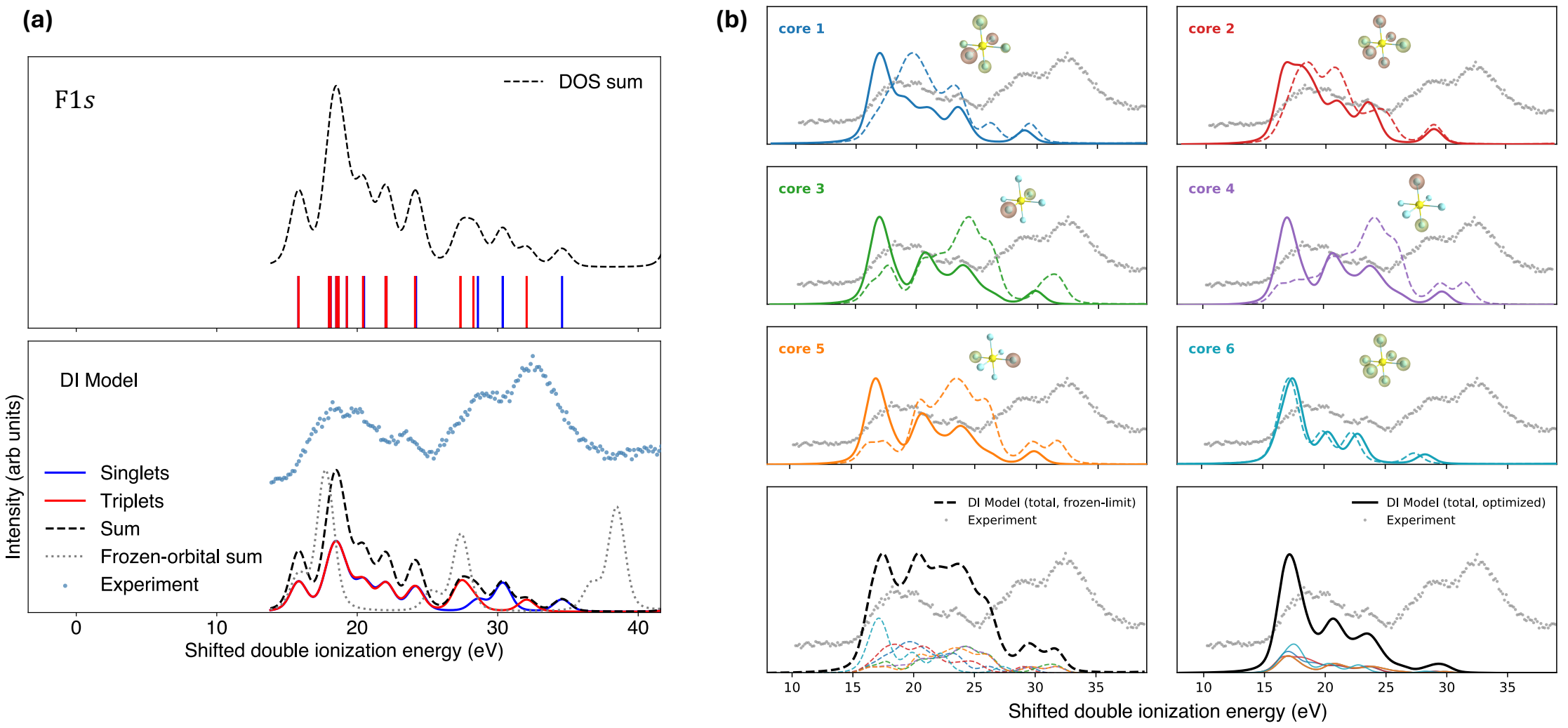

***(a)*** *The outer-shell CV states at the F1$s$ edge. (top) The bars indicate the energies of the CV states at the ΔSCF level, and the black dashed line represents the density of CV states (DOS). (bottom) The DI model (black dashed line) and the frozen-orbital limit (gray dotted line) as shown for comparison.* ***(b)*** *Calculations of CV spectra targeting the F1s orbitals in (a) using localized MOs, where $O_h$ symmetry breaking takes place, and in (b) using delocalized MOs, where the symmetry is preserved and assignments are made according to $O_h$ point group.*

Comparing the core-valence spectra with the valence photoelectron spectra, one can notice some differences, apart from the singlet-triplet splittings (analyzed in Fig. 2 **(a)**, and more thoroughly in Figs. S3 and S4 of the Supplementary Materials). In particular, one finds that the spectra are energetically stretched out in the core-valence cases. The S1$s$ and S2$p$ outer-valence spectra reach about 4 eV further down in energy than the 21 eV valence photoelectron spectrum. As argued earlier(Olsson, Cornetta, et al. 2025) the wider spread of the core-valence spectra arises from the deeper potential generated by the additional core hole. We note that the widening is about the same for the F1$s$ spectrum, although a different core site and presumably different core-hole screening is involved. The two sulphur spectra, S1$s$ and S2$p$, show almost identical features, except for the obvious lifetime broadening of the former and the different singlet-triplet (spin-orbit) splittings analyzed in Figs. 3 **(a)** and 3 **(b)**. The S2$p$ spectrum is split by spin-orbit coupling into two parallel progressions separated by 1.2 eV. The separation is thus a relativistic effect that, in principle, also generates a slight deviation from the 3:1 intensity ratio of the two $J$ components (1/2 and 3/2) in the conventional core-level photoelectron spectrum, and, as expected, also in the CV spectra. This deviation is, however, not resolvable under present conditions.

***(a)*** *The outer-shell CV states at the S1$s$ edge. (top) The bars indicate the energies of the CV states at the ΔSCF level, and the black dashed line represents the density of CV states*

*(DOS). (bottom) The DI model (black dashed line) and the frozen-orbital limit are shown for comparison.* ***(b)*** *The outer-shell CV states at the S2$p$ edge. (top) The bars indicate the energies of the CV states at the ΔSCF level, and the black dashed line represents the density of CV states (DOS). (bottom) The DI model (black dashed line) and the frozen-orbital limit (gray dotted line) are shown for comparison.*

Concerning singlet – triplet separations of the core-valence states, it is relevant to compare inner and outer MO levels and also how the splittings compare between the spectra. As recognized earlier for other systems (Olsson, Cornetta, et al. 2025), the S-T splittings in the CV spectra depend on the localization and binding energy of the molecular orbital involved. As we see in Figs. 2 **(a)** and 3, the splittings are very small, indeed negligible, for the outer MOs, while they increase towards higher binding energies. It is instructive here to compare the S1$s$ (Table 1) and S2$p$ (Table S1 in LS coupling) spectra, as they refer to the same core site. The splittings are in general larger for S2$p$; for example, for the 5a$_{1g}$ MO level at about 30 eV. This is due to the fact that the valence-to-core penetration is more effective for the shallower S2$p$ level than for the deeper S1$s$ level, creating a larger core-valence exchange interaction and a larger splitting between the singlet and triplet states. For the F1$s$ spectra, we also note a trend from almost zero splitting for the outer lying 1t$_{1g}$, 1t$_{2g}$, and 5t$_{1u}$ orbitals to a 2.5 eV splitting for the inner-lying 5a$_{1g}$ orbital. For the very deep-lying F2$s$-based orbitals at 45-50 eV, the predicted splitting is very large, but probably unobservable due to the strong MO breakdown effects and high background.

The S1$s$ and S2$p$ spectra, on the one hand, and the F1$s$ spectrum, on the other, also differ with respect to symmetry breaking and core-hole localization. Thus, while F1$s$ ionization breaks molecular symmetry, enforcing localized core holes through vibronic interaction between symmetry-adapted F1$s$ core-hole states and antisymmetric vibrational modes, no such core-hole-localizing antisymmetric vibronic coupling occurs following ionization of the central sulphur core. We have nevertheless assigned the fluorine CV spectrum also according to the high-symmetry labels of the ground-state orbitals to facilitate comparisons. These symmetry-breaking aspects for CV spectra have been discussed previously in Refs. (Ideböhn et al. 2023; Olsson, Cornetta, et al. 2025). As noted above, at higher ionization energies involving inner-valence levels, electron-correlation (MO breakdown) effects enter and are prevalent in both conventional valence photoelectron and CV spectra in general. For the latter, such effects may come into play also for the somewhat deeper outer levels, such as 5a$_{1g}$ in $SF_6$. This is due to the stronger CV potential, which generates more possibilities for near-degeneracies.

In the 30 – 40 eV region of the F1$s$ spectrum, we observe high intensity that is quite significantly underestimated by the OSRHF scheme. On the other hand, the frozen-orbital representation does provide intensity in this region, provided that we make a correction for the relaxation energy of the frozen orbital band at about 40 eV. We speculate that this originates from MO breakdown effects that are prevalent in this region (as also indicated by point RASSCF calculations). Considering the opening of a core hole, the sheer number of electrons and, in contrast to the S1$s$ and S2$p$ spectra, the lowering of symmetry of the F1$s$ hole states imply plentiful opportunities for near-degenerate CI excitations causing breakdown. As suggested in previous works(Ideböhn et al. 2023; Olsson, Cornetta, et al. 2025), this occurs earlier in core-valence spectra than for 2s levels in conventional valence photoelectron spectra, but also bears some reminiscence of the strong MO breakdown in larger parts of molecular Auger spectra, representing double-valence-hole states(Ågren 1981).

# Conclusions

Core-valence spectra at the S1$s$, S2$p$, and F1$s$ edges of $SF_6$ have been recorded and analyzed in this work with the help of quantum chemical calculations, revealing a multitude of well-separated structures covering wide energy intervals. As a novel theoretical contribution to CV spectroscopy, an intensity model for core-valence double ionization has been derived. It includes photoionization, dipolar coupling and shake-off, and monopolar coupling to the continuum. The model maintains proper spin coupling between the discrete doubly ionized states and the two continuum electrons. The application of this model, together with open-shell restricted Hartree-Fock optimization of the doubly-ionized state energies, made a detailed analysis of the experimental spectral features of the $SF_6$ CV spectra possible. It was found that assignments in terms of molecular orbitals can be applied over large intervals of the spectra, provided that the appropriate singlet-triplet spin separation of the dicationic states is considered.

Core ionizations of the central sulphur atom, both through S1$s$ and S2$p$, produce core-valence spectra that agree well with the well-known single-ionization spectrum. For the deep-lying S1$s$ MO, the measured spectrum has been successfully replicated using this simple DI model, see Fig. 3 **(a)**. The core-valence spectrum above S2$p$, with a much higher SO splitting than at the 1s sites, instead requires a different coupling regime in which $j$, rather than spin alone, is taken into consideration. Both spectra extend to deeper relative energies than the single-ionization spectrum, as expected from the deeper potential caused by the core hole.

In analyzing the F1$s$ core-valence ionization, the difference in appearance, both compared to the other core-valence spectra and to the single-ionization spectrum, is apparent. The first band of peaks covers the energy range of the outer-valence photoelectron peaks, but the latter peaks above 25 eV have a vastly different intensity profile compared to the other spectra. Calculations suggest that they are predominantly due to the innermost valence orbital $5a_{1g}^2$, and the intensities may be explained by MO breakdown effects in this region. Here, it is important to point out that this core-valence spectrum results from symmetry breaking of the MOs due to the core-hole localization, which may also explain the differences between the spectra.

# Funding

This work has been financially supported by the Swedish Research Council (grant number 2023-03464) and the Knut and Alice Wallenberg Foundation (grant numbers 2024.0120), Sweden. F.T. acknowledges funding by the Deutsche Forschungsgemeinschaft (DFG, German Research Foundation) - Project 509471550, Emmy Noether Programme.

# Acknowledgment

We acknowledge DESY (Hamburg, Germany), a member of the Helmholtz Association HGF, for the provision of the experimental facilities. Parts of this research were carried out at PETRA III. Data was collected using beamline P04 operated by DESY Photon Science. We would like to thank Jörn Seltmann and Moritz Hoesch for assistance during the experiments. Beamtime was allocated for proposal I-20230132. Special thanks go to Dr. Valerio Bassetti from the MAX-IV laboratory in Lund for the development of the vital synchronization system of the mechanical chopper used in this work.

# Figure legends

**Figure 1** Core-valence double-ionization electron-pair spectra of $SF_6$ obtained at photon energies well above the core S2$p$ , F1s, and S1$s$ energies at 270, 797, and 2599 eV, respectively, in comparison to the well-known single-ionization valence electron spectra obtained at 21 eV (lowest panel), and 41 eV (highest panel). The error bars in the plots reflect the statistical uncertainty, except for the valence electron spectra, which are plotted without error bars owing to the high number of counts obtained.
**Figure 2(a)** The outer-shell CV states at the F1$s$ edge. (top) The bars indicate the energies of the CV states at the ΔSCF level, and the black dashed line represents the density of CV states (DOS). (bottom) The DI model (black dashed line) and the frozen-orbital limit (gray dotted line) as shown for comparison. **(b)** Calculations of CV spectra targeting the F1$s$ orbitals in (a) using localized MOs, where $O_h$ symmetry breaking takes place, and in (b) using delocalized MOs, where the symmetry is preserved and assignments are made according to $O_h$ point group.
**Figure 3 (a)** The outer-shell CV states at the S1$s$ edge. (top) The bars indicate the energies of the CV states at the ΔSCF level, and the black dashed line represents the density of CV states (DOS). (bottom) The DI model (black dashed line) and the frozen-orbital limit are shown for comparison. **(b)** The outer-shell CV states at the S2$p$ edge. (top) The bars indicate the energies of the CV states at the ΔSCF level, and the black dashed line represents the density of CV states (DOS). (bottom) The DI model (black dashed line) and the frozen-orbital limit (gray dotted line) are shown for comparison.

# Tables

*Per-state DI integrals for the CV channels over the S1$s$ edge. The orbital indices denote $i$ = core hole and $j$ = valence hole, so $\mathcal{D}_i\mathcal{S}_j$ couples the core dipole amplitude to the valence shake-off, while $\mathcal{D}_j\mathcal{S}_i$ does the opposite; $\mathcal{G}_{ij}$ is the $\alpha\alpha/\beta\beta$ dication-orbital term, which is absent in the singlet states. The "State" column gives the $O_h$ irrep of the valence hole, assigned by projecting the dication valence-hole MO onto the neutral-molecule shells. Energies are in eV relative to the lowest CV state, and all integrals are given in atomic units $\times 10^3$.*

| State | Spin (dic.) | $E_{\rm rel}$ (eV) | $\int \mathcal{D}_j\,\mathcal{S}_i$ | $\int \mathcal{D}_i\,\mathcal{S}_j$ | $\int \mathcal{G}_{ij}$ |
|---|---|---|---|---|---|
| $3e_g$ | T | 0.00 | 0.03 | 2.03 | 2.05 |
| $3e_g$ | S | 0.00 | 0.03 | 2.03 | – |
| $1t_{1g}$ | T | 0.12 | 0.04 | 2.69 | 2.74 |
| $1t_{1g}$ | S | 0.12 | 0.04 | 2.69 | – |
| $5t_{1u}$ | T | 1.15 | 0.04 | 2.59 | 2.63 |
| $5t_{1u}$ | S | 1.17 | 0.06 | 2.59 | – |
| $1t_{2u}$ | T | 1.40 | 0.03 | 2.55 | 2.58 |
| $1t_{2u}$ | S | 1.40 | 0.03 | 2.55 | – |
| $1t_{2g}$ | T | 4.15 | 0.03 | 2.23 | 2.26 |
| $1t_{2g}$ | S | 4.15 | 0.03 | 2.23 | – |
| $4t_{1u}$ | T | 7.93 | 0.06 | 1.97 | 2.03 |
| $4t_{1u}$ | S | 8.12 | 0.06 | 1.98 | – |
| $5a_{1g}$ | T | 14.71 | 0.01 | 1.47 | 1.60 |
| $5a_{1g}$ | S | 15.39 | 0.01 | 1.50 | – |

*Per-state DI integrals for the CV channels over the S2$p$ edge. The orbital indices denote $i$ = core hole and $j$ = valence hole, so $\mathcal{D}_i\mathcal{S}_j$ couples the core dipole amplitude to the valence shake-off, while $\mathcal{D}_j\mathcal{S}_i$ does the opposite; $\mathcal{G}_{ij}$ is the $\alpha\alpha/\beta\beta$ dication-orbital term, which is absent in the singlet states. Energies are in eV relative to the lowest CV state, and all integrals and intensities are given in atomic units $\times 10^3$.*

| State | $j_C$ | $E_{\rm rel}$ | $\int \mathcal{D}_j\,\mathcal{S}_i$ | $\int \mathcal{D}_i\,\mathcal{S}_j$ | $\int \mathcal{G}_{ij}$ | $I^{S_{\rm dic}=0}$ | $I^{S_{\rm dic}=1}$ | $I^{(j_C)}$ |
|---|---|---|---|---|---|---|---|---|
| $1t_{1g}$ | 3/2 | 0.00 | 1.71 | 4.60 | 6.31 | 39.7 | 39.8 | 79.6 |
| | 1/2 | 1.12 | 1.72 | 4.72 | 6.43 | 40.5 | 40.5 | 40.6 |
| $3e_g$ | 3/2 | 0.24 | 1.37 | 3.49 | 4.86 | 24.4 | 23.8 | 48.2 |

| State | $j_C$ | $E_{\rm rel}$ | $\int \mathcal{D}_j\,\mathcal{S}_i$ | $\int \mathcal{D}_i\,\mathcal{S}_j$ | $\int \mathcal{G}_{ij}$ | $I^{S_{\rm dic}=0}$ | $I^{S_{\rm dic}=1}$ | $I^{(j_C)}$ |
|---|---|---|---|---|---|---|---|---|
| | 1/2 | 1.36 | 1.38 | 3.57 | 4.95 | 24.9 | 24.3 | 24.5 |
| $5t_{1u}$ | 3/2 | 1.00 | 2.02 | 4.16 | 6.18 | 38.2 | 37.4 | 75.7 |
| | 1/2 | 2.13 | 2.03 | 4.26 | 6.28 | 38.9 | 38.1 | 38.5 |
| $1t_{2u}$ | 3/2 | 1.28 | 1.58 | 4.07 | 5.65 | 39.6 | 38.7 | 78.2 |
| | 1/2 | 2.40 | 1.58 | 4.16 | 5.75 | 40.3 | 39.3 | 39.9 |
| $1t_{2g}$ | 3/2 | 4.08 | 1.31 | 3.83 | 5.14 | 42.1 | 41.8 | 84.0 |
| | 1/2 | 5.20 | 1.31 | 3.93 | 5.24 | 42.8 | 42.6 | 42.7 |
| $4t_{1u}$ | 3/2 | 7.59 | 2.96 | 3.18 | 6.13 | 36.5 | 36.9 | 73.4 |
| | 1/2 | 8.71 | 2.97 | 3.25 | 6.21 | 37.1 | 37.2 | 37.2 |
| $5a_{1g}$ | 3/2 | 14.36 | 5.95 | 2.37 | 8.32 | 16.7 | 16.6 | 33.4 |
| | 1/2 | 15.70 | 5.98 | 2.43 | 8.40 | 16.9 | 16.8 | 16.8 |

*Per-state DI integrals for the CV channels over the F*$1s$ *edge. The orbital indices denote* $i =$ *core hole and* $j =$ *valence hole, so* $\mathcal{D}_i\mathcal{S}_j$ *couples the core dipole amplitude to the valence shake-off, while* $\mathcal{D}_j\mathcal{S}_i$ *does the opposite;* $\mathcal{G}_{ij}$ *is the* $\alpha\alpha/\beta\beta$ *dication-orbital term, which is absent in the singlet states. The "State" column gives the* $O_h$ *irrep of the valence hole, assigned by projecting the dication valence-hole MO onto the neutral-molecule shells. Energies are in eV relative to the lowest CV state, and all integrals are given in atomic units* $\times 10^3$.

| Spin (dic.) | Spin | $E_{\rm rel}$ (eV) | $\int \mathcal{D}_j\,\mathcal{S}_i$ | $\int \mathcal{D}_i\,\mathcal{S}_j$ | $\int \mathcal{G}_{ij}$ |
|---|---|---|---|---|---|
| $1t_{1g} + 1t_{2u} + 5t_{1u}$ | T | 0.00 | 0.18 | 3.13 | 3.31 |
| | S | 0.00 | 0.18 | 3.13 | – |
| | T | 0.00 | 0.12 | 3.13 | 3.25 |
| | S | 0.00 | 0.12 | 3.13 | – |
| $1t_{1g}$ | T | 2.13 | 0.12 | 3.35 | 3.47 |
| | S | 2.13 | 0.12 | 3.35 | – |

| Spin (dic.) | Spin | $E_{\rm rel}$ (eV) | $\int \mathcal{D}_j \, \mathcal{S}_i$ | $\int \mathcal{D}_i \, \mathcal{S}_j$ | $\int \mathcal{G}_{ij}$ |
|---|---|---|---|---|---|
| $1t_{2g} + 1t_{2u} + 1t_{1g}$ | T | 2.17 | 0.15 | 3.03 | 3.17 |
| | S | 2.18 | 0.15 | 3.03 | – |
| $3t_{2g} + 1t_{2u}$ | T | 2.69 | 0.13 | 2.79 | 2.92 |
| | S | 2.69 | 0.13 | 2.79 | – |
| $1t_{2u} + 5t_{1g}$ | T | 2.84 | 0.21 | 3.16 | 3.37 |
| | S | 2.84 | 0.21 | 3.16 | – |
| | T | 2.84 | 0.14 | 3.16 | 3.30 |
| | S | 2.84 | 0.14 | 3.16 | – |
| $1t_{2u} + 3t_{2g}$ | T | 3.44 | 0.12 | 3.19 | 3.31 |
| | S | 3.44 | 0.12 | 3.19 | – |
| $1t_{2g} + 1t_{1g} + 1t_{2u}$ | T | 4.60 | 0.10 | 2.97 | 3.08 |
| | S | 4.66 | 0.10 | 2.97 | – |
| | T | 4.60 | 0.12 | 2.97 | 3.10 |
| | S | 4.66 | 0.12 | 2.97 | – |
| $1t_{2g}$ | T | 6.20 | 0.13 | 2.85 | 2.98 |
| | S | 6.20 | 0.13 | 2.85 | – |
| $4t_{1u} + 3t_{2g} + 5t_{1u}$ | T | 6.02 | 0.12 | 2.85 | 2.97 |
| | S | 6.28 | 0.12 | 2.86 | – |
| $4t_{1g}$ | T | 8.30 | 0.26 | 2.51 | 2.77 |
| | S | 8.36 | 0.26 | 2.51 | – |
| | T | 8.30 | 0.27 | 2.51 | 2.78 |
| | S | 8.36 | 0.27 | 2.51 | – |
| $1t_{2g} + 1t_{2u} + 1t_{1u}$ | T | 11.53 | 0.12 | 2.45 | 2.57 |
| | S | 14.54 | 0.12 | 2.61 | – |
| | T | 11.53 | 0.10 | 2.45 | 2.55 |
| | S | 14.54 | 0.10 | 2.61 | – |
| $5a_{1g} + 4t_{1u} + 3t_{2g}$ | T | 12.34 | 0.42 | 2.18 | 2.60 |
| | S | 12.77 | 0.43 | 2.20 | – |
| $5a_{1g} + 4t_{1u} + 3t_{2g}$ | T | 16.24 | 0.29 | 2.06 | 2.35 |
| | S | 18.76 | 0.29 | 2.17 | – |

*List of peak assignments in the single-ionization-energy (SIE) spectra compared to literature values and the shifted core-valence spectra. The literature values refer to highest-intensity vibrational peaks when vibrational structure is resolved.*

| | List of peak assignments in the single-ionization-energy (SIE) spectra compared to literature values and the shifted core-valence spectra. The literature values refer to highest-intensity vibrational peaks when vibrational structure is resolved. | *List of peak assignments in the single-ionization-energy (SIE) spectra compared to literature values and the shifted core-valence spectra. The literature values refer to highest-intensity vibrational peaks when vibrational structure is resolved.* | *List of peak assignments in the single-ionization-energy (SIE) spectra compared to literature values and the shifted core-valence spectra. The literature values refer to highest-intensity vibrational peaks when vibrational structure is resolved.* |
|---|---|---|---|
| Assignment | Lit. values(Holland et al. 1995) (eV) | SIE, $h\nu$ = 21 eV (eV) | SIE, $h\nu$ = 41 eV (eV) |
| $X\ ^2T_{1g}$ | 15.67 | 15.65 | 15.7 |
| $A\ ^2T_{1u}$ | 16.95 | 16.93 | 17.0 |
| $B\ ^2T_{2u}$ | 17.20 | 17.23 | |
| $C\ ^2E_g$ | 18.35 | 18.27 | 18.4 |
| $D\ ^2T_{2g}$ | 19.68 | 19.75 | 19.8 |
| $E\ ^2T_{1u}$ | 22.53 | | 22.5 |
| $F\ ^2A_{1g}$ | 26.82 | | 26.8 |

Resolving power in the 21 eV SIE is approximately 0.05 eV for the first peak and 0.01 eV for the last peak.

The resolving power is about 0.2 eV for the first peak and 0.12 eV for the last peak listed.

*List of energies from the shifted core-valence spectra. The values in the core-valence (CV) columns are those visible in Fig. 1, and are shifted by -187.35 eV, -700.7 eV, and -2497.6 eV, respectively for the S2$p$,F1$s$, and S1$s$ measurements. The bands in the spectra have been assigned to values calculated by us and listed in Tables 2,3,1 based on the best match.*

| Assignment | CV, Above S2$p$ | Assignment | CV, Above F1$s$ | Assignment | CV, Above S1$s$ |
|---|---|---|---|---|---|
| $1t_{1g}(3/2)$ | 15.7 | $1t_{1g}$+$1t_{2u}$+$5t_{1u}$ | 15.4 | $3e_g$+$1t_{1g}$ | 15.9 |
| $3e_g(1/2)$+$1t_{2u}(3/2)$+5 | 17.0 | $1t_{1g}$/$1t_{1g}$+$1t_{2g}$+$1t_{2u}$ | 17.3 | $1t_{2u}$+$5t_{1u}$ | 17.6 |

List of energies from the shifted core-valence spectra. The values in the core-valence (CV) columns are those visible in Fig. 1, and are shifted by -187.35 eV, -700.7 eV, and -2497.6 eV, respectively for the S2$p$,F1$s$, and S1$s$ measurements. The bands in the spectra have been assigned to values calculated by us and listed in Tables 2,3,1 based on the best match.

| Assignment | CV, Above S2$p$ | Assignment | CV, Above F1$s$ | Assignment | CV, Above S1$s$ |
|---|---|---|---|---|---|
| $t_{1u}(3/2)$ | | | | | |
| 1$t_{2u}(1/2)$+5 $t_{1u}(1/2)$ | 18.4 | 3$t_{2g}$+1$t_{2u}$ | 19.0 | 1$t_{2g}$ | 20.6 |
| 1$t_{2g}(3/2)$ | 19.7 | 1$t_{2g}$ | 21.5 | | |
| | | 4$t_{1g}$ | 22.4 | | |

List of energies from the shifted core-valence spectra. The values in the core-valence (CV) columns are those visible in Fig. 1, and are shifted by -187.35 eV, -700.7 eV, and -2497.6 eV, respectively for the $S2p$,$F1s$, and $S1s$ measurements. The bands in the spectra have been assigned to values calculated by us and listed in Tables 2,3,1 based on the best match.

| Assignment | CV, Above $S2p$ | Assignment | CV, Above $F1s$ | Assignment | CV, Above $S1s$ |
|---|---|---|---|---|---|
| $4t_{1u}(3/2)$ | 23.3 | | | | |
| $4t_{1u}(1/2)$ | 24.9 | $5a_{1g}$+$4t_{1u}$+$3t_{2g}$ | 28.7 | $4t_{1u}$ | 24.6 |
| $5a_{1g}(3/2)$ | 29.6 | | | $5a_{1g}$ | 29.4 |
| | | $5a_{1g}$+$4t_{1u}$+$3t_{2g}$ | 31.4 | $5a_{1g}$ | 31.7 |
| | | $5a_{1g}$+$4t_{1u}$+$3t_{2g}$ | 32.0 | | |

The lowest resolving power for the first listed peak is about 1.3 eV.

The lowest resolving power for the first listed peak is about 1.6 eV.

The lowest resolving power for the first listed peak is about 1.9 eV.